# Atomic and electronic structure of a copper/graphene interface as prepared and 1.5 years after

D. W. Boukhvalov[1,2*], P. F. Bazylewski[3], A.I. Kukharenko[4,5], I. S. Zhidkov[5], Yu. S. Ponosov[4], E.Z. Kurmaev[4,5], S. O. Cholakh[5], Y. H. Lee[6,7] and G. S. Chang[3]

[1]Department of Chemistry, Hanyang University, 17 Haengdang-dong, Seongdong-gu, Seoul 04763, Korea

[2]Theoretical Physics and Applied Mathematics Department, Ural Federal University, Mira Street 19, 620002 Yekaterinburg, Russia

[3]Department of Physics and Engineering Physics, University of Saskatchewan, 116 Science Place Saskatoon, Canada

[4]M.N. Mikheev Institute of Metal Physics, Russian Academy of Sciences-Ural Division, 620990 Yekaterinburg, Russia

[5]Institute of Physics and Technology, Ural Federal University, 9 Mira Street, 620002 Yekaterinburg, Russia

[6] Center for Integrated Nanostructure Physics, Institute for Basic Science, Sungkyunkwan University, Suwon 440-746, Korea

[7]Department of Physics and Department of Energy Science, Sungkyunkwan University, Suwon 440-746, Korea

*We report the results of X-ray spectroscopy and Raman measurements of as-prepared graphene on a high quality copper surface and the same materials after 1.5 years under different conditions (ambient and low humidity). The obtained results were compared with density functional theory calculations of the formation energies and electronic structures of various structural defects in graphene/Cu interfaces. For evaluation of the stability of the carbon cover, we propose a two-step model. The first step is oxidation of the graphene, and the second is perforation of graphene with the removal of carbon atoms as part of the carbon dioxide molecule. Results of the modeling and experimental measurements provide evidence that graphene grown on high-quality copper substrate becomes robust and stable in time (1.5 years). However, the stability of this interface depends on the quality of the graphene and the number of native defects in the graphene and substrate. The effect of the presence of a metallic substrate with defects on the stability and electronic structure of graphene is also discussed.*

# 1. Introduction

Since the earliest reports of formation of graphene-metal interfaces [1–3], this subject was discussed in multiple experimental [1–20] and theoretical papers [19–25]. There are four reasons that have motivated these studies. The first is usage of the metallic substrate for the growth of graphene [3, 13–17]. The second is protection of metallic surfaces from fast and unavoidable oxidation [1,5–11,23-25.]. The third is fabrication of graphene-metal contacts for various devices [2,4,19,21], and the fourth is application of graphene-metal composites as catalysts [18,19,25]. Additional motivation for the studying of graphene/metal interfaces is the production of biocompatible stable nanoparticles [27,28] and its further functionalization for tearapeutic and drug delivery applications. However, further developments in the area of graphene-metal interfaces require understanding the influences of the quality of metallic substrate on the electronic structure of graphene and its stability.

Recent experimental work has considered the effect of high-quality monocrystalline metallic substrate on the electronic structure of graphene [5,6]. However, the influence of imperfections on metallic substrates such as grain boundaries and various impurities on the electronic structure and stability of graphene has not been closely examined. One of the main routes to explore the role of the substrate on the stability of the carbon cover is long-term studies of changes in the atomic and electronic structure of these materials. Several studies have demonstrated the stability of either graphene or metallic substrate after long-term exploration in ambient conditions and annealing [7]. Other experimental work has revealed the presence of traces of an oxide phase in metallic substrates [10]. Changes in the wettability of metals supporting graphene after annealing [11] and oxidation of chemically doped graphene [12] have also been demonstrated. However, graphene has been found to be suitable only for short-term protection of copper substrate [8]. Note that the scanning electron microscopy images of Cu-graphene interfaces with non-oxidized [5] or oxidized [7] metallic substrate are rather similar. These results demonstrate that the factors governing the stability of the graphene/metal interface remain unclear.

From the theoretical side, the mechanism of decay of the graphene/metal interface also remains unclear. There are two possible paths for the penetration of oxygen to a metal surface covered with graphene. The first is intercalation where gas molecules leak between the graphene and metal to interact with the host metal. This model explains why oxidation requires heating during the intercalation process. However, it explains neither the dependence of the metal corrosion on the quality of graphene (which should not affect the process of intercalation) nor why corrosion starts from the sample edges instead of the center [9]. The second model describes penetration of oxygen through the surface of graphene [8], but according to theory and experiments, graphene is impermeable to gasses, even atomic helium [29,30]. The energy barrier for this penetration is also rather high (about 3–5 eV) [25,26], which cannot explain the oxidation of the substrate at ambient conditions reported in some experimental work [31]. On the other hand, we have shown that doped graphene may be permeable due to the formation of defects, which serve as permeation pathways [32]. Recent experiments have demonstrated the possibility of the formation of vacancies in one, two, and three-layer graphene in the process of oxidation [33]. A model of interaction between oxygen and graphene on copper was proposed, taking into account that some carbon atoms can penetrate into the metal substrate during synthesis of the graphene on the metal surface. One can assume that the introduction of carbon into the metal can be realized at grain boundaries, which are inevitably present in the copper foil [13].

In the present paper, we performed density functional theory (DFT) calculations with the SIESTA software package of a graphene/Cu interface with graphene both intact and containing vacancies on copper, as well as with carbon atoms as Cu-interstitials. The obtained results are compared with extresurface-sensitive X-ray photoelectron spectra (XPS) and X-ray absorption near the edge structure (XANES) spectra, which are used for characterization of graphene-coated copper after exposure to ambient air from 1 month to 1.5 years. Graphene was grown on copper substrates using chemical vapor deposition (see Ref. [21] for details). Samples were stored at both a low humidity (LH) environment in a desiccator and at room temperature (RT).

## 2. Experimental methods

The spectroscopic characterization of graphene/Cu composites begins with XPS survey spectra of the elements in the samples, as presented in Fig. 1(a). XPS spectra were measured using a PHI XPS Versaprobe 5000 spectrometer (ULVAC-Physical Electronics, USA) with Al *Kα* excitation at 1486.6 eV with a resolution of $\Delta E \leq 0.5$ eV. The core level peaks associated with carbon (C 1*s*) and copper (Cu 2*s*, Cu 2*p*, Cu 3*s*, Cu 3*p* and Cu LMM) are clearly seen in each survey spectrum. A small amount of oxygen (O 1*s*) is detected but did not vary between the samples. The inset of Fig. 1(a) displays high-resolution measurements of the C 1*s* portion showing that the graphene film is continuous and very close to a full monolayer in coverage. The low-intensity feature at 288.4 eV corresponds to light oxidation of the graphene. The observation of only a trace amount of oxygen indicates that the graphene overlayer is inert and resists oxidation and oxygen diffusion into the metal substrate. Even after 1.5 years of exposure, the oxidation is comparable to freshly prepared samples. The Raman spectra in Fig. 1(b) show the broad intense luminescence peak typical for fresh electropolished copper with two spikes from the graphene *G* and 2*D* bands which appear only in the spectra of the graphene/Cu composites. The Raman spectra were excited with a 532 nm (2.33 eV) laser with an average power of 3 mW focused onto the sample by a 50 times objective lens with a numerical aperture of 0.8. The *G* and 2*D* bands do not vary significantly in either intensity or width between the samples, indicating that the graphene is not affected by aging compared to freshly prepared graphene. Cu-foil oxidized in ambient air shows a shift towards a higher frequency of ~1000 cm$^{-1}$ that has not been observed for graphene/Cu composites.

## 3. Computational methods

To explain the peak splitting observed in the C 1*s* XANES results, we first consider the influence of carbon defects on the electronic structure of graphene on Cu substrate. For DFT calculations of graphene systems, the pseudopotential code SIESTA was used [34]. For specific details, refer to other studies [19,20,25,26,29]. Calculations were performed using local density approximation [35], which is feasible for modeling graphene over a copper substrate because formation of the chemical not the van der Waals bonds occurs between the 3d orbitals of the metals and the π-orbitals of graphene. This issue was discussed in detail in Ref. 20. A full optimization of the atomic positions was performed, as well as optimization of the force and total energy with an accuracy of 0.04 eV/Å and 1 meV, respectively. The wave functions were expanded with a double-ζ plus polarization basis of localized orbitals for carbon and oxygen. All calculations were carried out with an energy mesh cut-off of 360 Ry and a **k**-point mesh of 4×4×2 in the Monkhorst-Pack scheme [36]. Calculation of the chemisorption energies for oxidation was performed using a standard equation: $E_{chem} = (E_{host+O} - [E_{host} + E_{O2}/2])$ where $E_{host}$ is the total energy of the system before adsorption of an oxygen atom and $E_{O2}$ is the total energy of the molecular oxygen in gas phase in the triplet state.

## 4. Experimental Results

The protection of the Cu surface with a graphene coating was also verified as shown by the Cu $2p_{3/2,1/2}$ core level XPS spectra presented in Fig. 2. Both graphene/Cu and Cu-metal have nearly the same spectral shape and energy position of Cu $2p_{3/2}$ and $2p_{1/2}$ core-excitation peaks [15] whereas the reference spectrum of the Cu-foil (oxidized) exhibits a shift to higher energy towards the binding energies of CuO [34]. In addition, the higher energy fine structures, $S_1$ and $S_2$, which are in the energy location for the charge-transfer satellites typical of CuO, are small or not visible in the Cu foil, demonstrating that both graphene/Cu and Cu-foil have no $Cu^{2+}$ oxidation state.

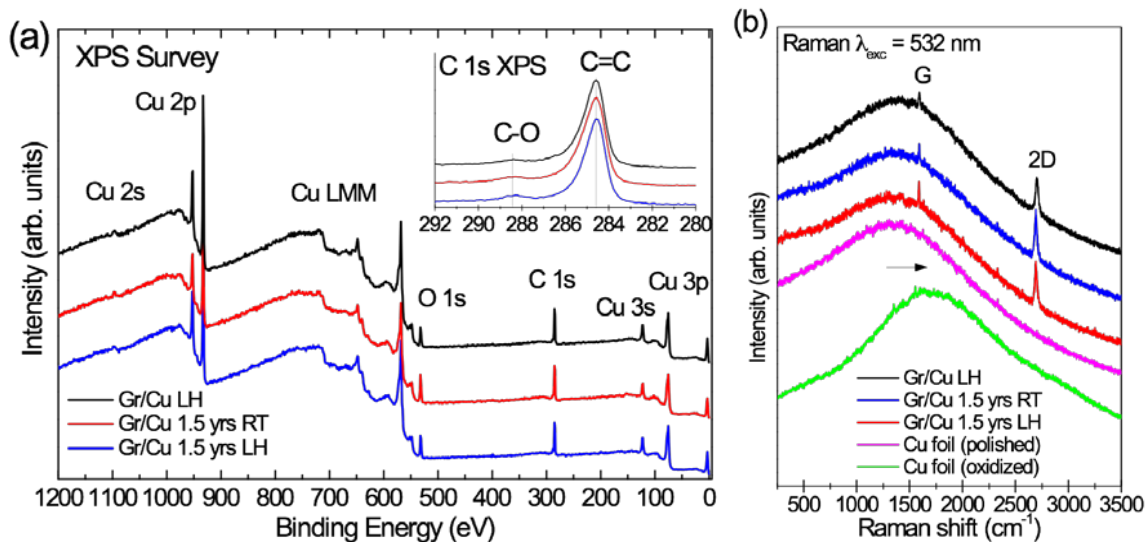

**Figure 1** (a). XPS survey spectra of graphene/copper (Gr/Cu) composites after oxygen exposure in low and high humidity environments with C 1*s* core level XPS spectra shown in the inset. (b) Raman spectra of graphene/Cu composites compared to the polished and oxidized Cu foil.

It should be noted, however, that the Cu $2p_{3/2,1/2}$ core-level XPS spectrum of Cu metal is identical to that of $Cu_2O$.[23] Therefore, the presence of a monovalent ($Cu^+$) oxidation state in the graphene/Cu composites cannot be excluded, and identification of the $Cu^+$ species is performed with the help of XPS VB spectra combined with XANES measurements, which show different near edge features for Cu, $Cu_2O$, and CuO.[23,24] As seen in Fig. 3, the XPS

VB spectra show the *A, B, C,* and *D* fine structure features typical for pure Cu metal, [25] which are completely reproduced in the spectra of graphene/Cu composites. Peaks *a* and *b* of the Cu-foil after exposure to ambient air at room temperature appear closer to that of $Cu_2O$ than CuO; no low-energy shift of spectral features is observed, and additional fine structure features *c, d,* and *e* appear, typical for the VB of CuO [37]. It is clear that the higher energy fine structure of $Cu_2O$ and CuO (features *c, d,* and *e*) are not seen in the VB spectra of graphene/Cu, suggesting that the Cu substrate is not oxidized.

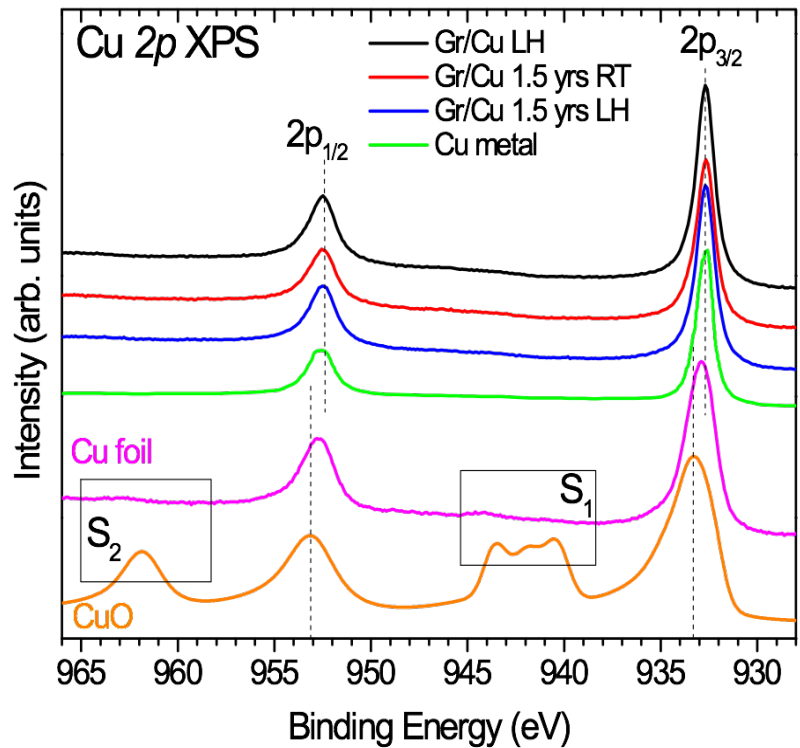

**Figure 2.** Comparison of the Cu $2p_{3/2,1/2}$ XPS spectra of the graphene/Cu composite with spectra of the reference samples (Cu-foil after exposure to ambient air, Cu metal, and CuO).

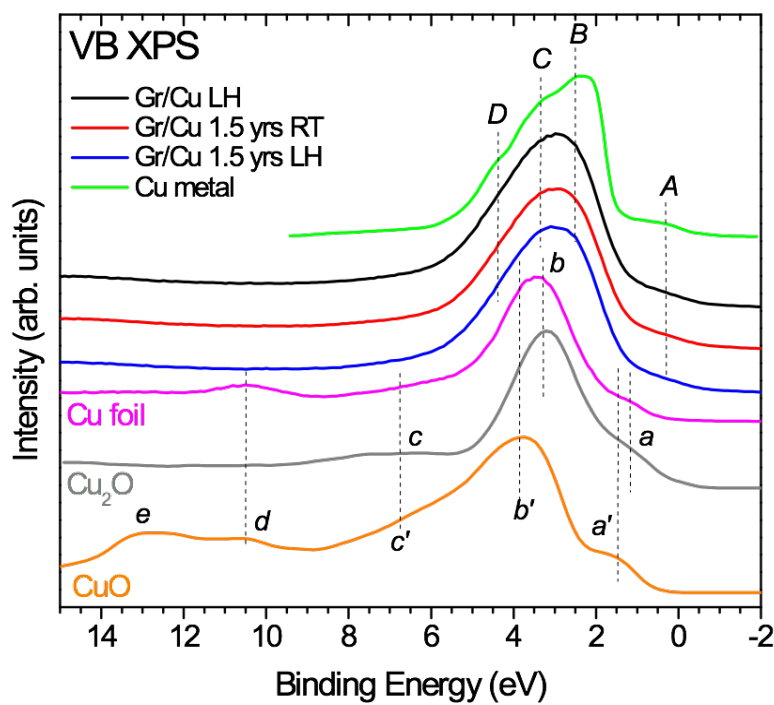

**Figure 3.** XPS valence band spectra of the graphene/Cu composite and the reference samples (Cu-foil after exposure to ambient air, Cu metal, CuO, and $Cu_2O$).

To further examine the electronic structure of graphene/Cu, the unoccupied density of the states (DOS) was probed by measuring the Cu 2$p$ and C 1$s$ XANES spectra (Fig. 4). XANES measurements were performed at the resonant inelastic and elastic scattering beamline of the Canadian Light Source (CLS), Canada. XANES spectra were measured in total electron yield mode with a resolution of 0.1 eV. The spectra were normalized to the incoming photon flux as recorded by the Au mesh, and the intensity was normalized to a constant background as follows: C 1$s$ XANES at 310 eV and Cu 2$p$ XANES at 990 eV. The graphene/Cu samples were compared to graphene/$SiO_2$, which was prepared by the standard chemical transfer method using poly(methyl methacrylate) (PMMA) [14]. Graphene/Cu shows splitting of the main π* peak at ~0.54 eV relative to graphene/$SiO_2$ with no change in the σ* peak position. This π* splitting suggests a change in the graphene electronic structure on Cu that can be attributed to the charge transfer, covalent bonding, and/or vacancy formation. A low concentration of C-O bonds is evident in all the graphene/Cu samples, but it is higher in the

graphene/SiO$_2$ transferred samples, likely due to residual solvent or PMMA from the transfer process.

Turning to Fig. 4 (b), it can be clearly seen from the comparison to the reference samples that the Cu 2*p* XANES spectra of the graphene/Cu samples are very similar to that of the Cu-metal in the energy position of the peaks as well as the fine structure, which is quite different from Cu$_2$O and CuO [38,39]. The Cu 2*p* XANES spectrum of the Cu foil (Alfa Aesar Lot No. 13380, 99.9%, 0.127 mm thickness) exposed to oxygen is by contrast found to be similar to Cu$_2$O with traces of CuO, in agreement with the XPS results. However, XANES measurements show that the Cu atoms on the surface of the Cu foil after long exposure at room temperature have a monovalent oxidation state (1+), inconsistent with the results found in Ref. 40.

In order to understand the observed high protective properties of graphene on Cu, it should be noted that, among the 3*d*-transition metals, copper has the lowest affinity to carbon and does not naturally form any carbide phases. Compared to Co and Ni, Cu has very low carbon solubility (0.001–0.008 weight % at 1084 °C) [15]. The low reactivity with carbon could be due to the fact that copper has the most stable electron configuration with a filled *d*-electron shell.

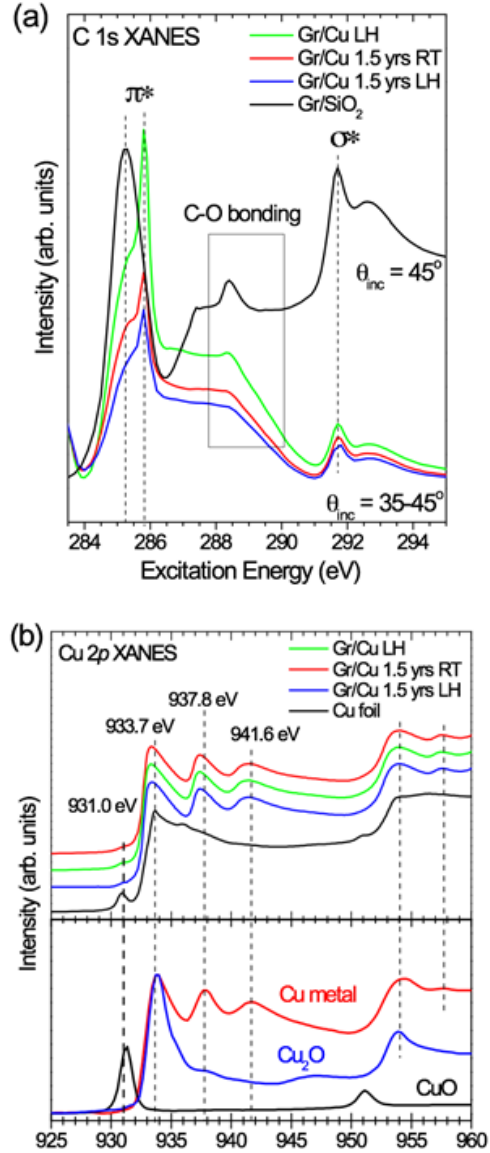

**Figure 4.** (a) C 1s XANES spectra of graphene/Cu composites compared to graphene/SiO$_2$, (b) Cu 2p XANES spectra of graphene/Cu compared to the Cu reference samples (bottom panel).

## 5. Computational results and comparison with experiments

The calculated density of the C 2p unoccupied states for free standing graphene is shown in Fig. 5(a). It is in reasonable agreement with the experimental C 1s XANES spectrum of the transferred graphene/SiO$_2$ sample and well reproduces the energy difference between the π and σ peaks (<7 eV). As seen in Fig. 4(a), C 1s XANES of Gr/Cu shows a 0.54 eV splitting, which has not been observed before in similar samples [31-43]. DFT calculations suggest that the presence of a copper substrate does not lead to great changes in the electronic structure of graphene (Fig. 5a,c), due to the weak coupling between graphene and the Cu substrate [21]. In order to explore the origin of the 0.54 eV splitting of the main π* peak, we checked the presence of bi-vacancies in graphene. This is a fairly stable configuration with a vacancy size that allows oxygen atoms to penetrate into the graphene and induce oxidation. The presence of this type of defect results in the formation of covalent bonds between the carbon atoms at the edges of the vacancies and the copper substrate. These covalent bonds lead to the appearance of an additional σ' peak and high energy shift in the π*-peak of 1.3 eV, as shown in Fig. 5(b). In contrast, they are not seen in the experimental C 1$s$ XANES spectra of the graphene/Cu samples. The presence of carbon defects in the substrate does not provide a visible influence on the electronic structure of the graphene cover (Fig. 5c). In some cases, oxidation of the Cu substrate for graphene/Cu composites has been experimentally observed [7,8]. This finding makes it necessary to estimate the possibility (see below) of this process and its effect on the electronic structure of graphene. However, Fig. 5(d) shows splitting of the π*-peak of <0.7 eV in the case where oxygen atoms form epoxy groups (C-O-C) combined with carbon impurities in the Cu matrix. This theoretical prediction is clearly confirmed in experimental C 1$s$ XANES spectra, which show splitting of the π*-peak of 0.54 eV, Fig. 4a). Based on a comparison of the experimental C 1$s$ XANES spectra with unoccupied C 2$p$ DOS calculated by DFT, we can fully exclude the formation of carbon vacancies over time as it would cause oxidation of the Cu. It is more likely that splitting of the σ*-peak results from sparse epoxide formation on the graphene surface combined with interstitial C impurity atoms in the metallic substrate.

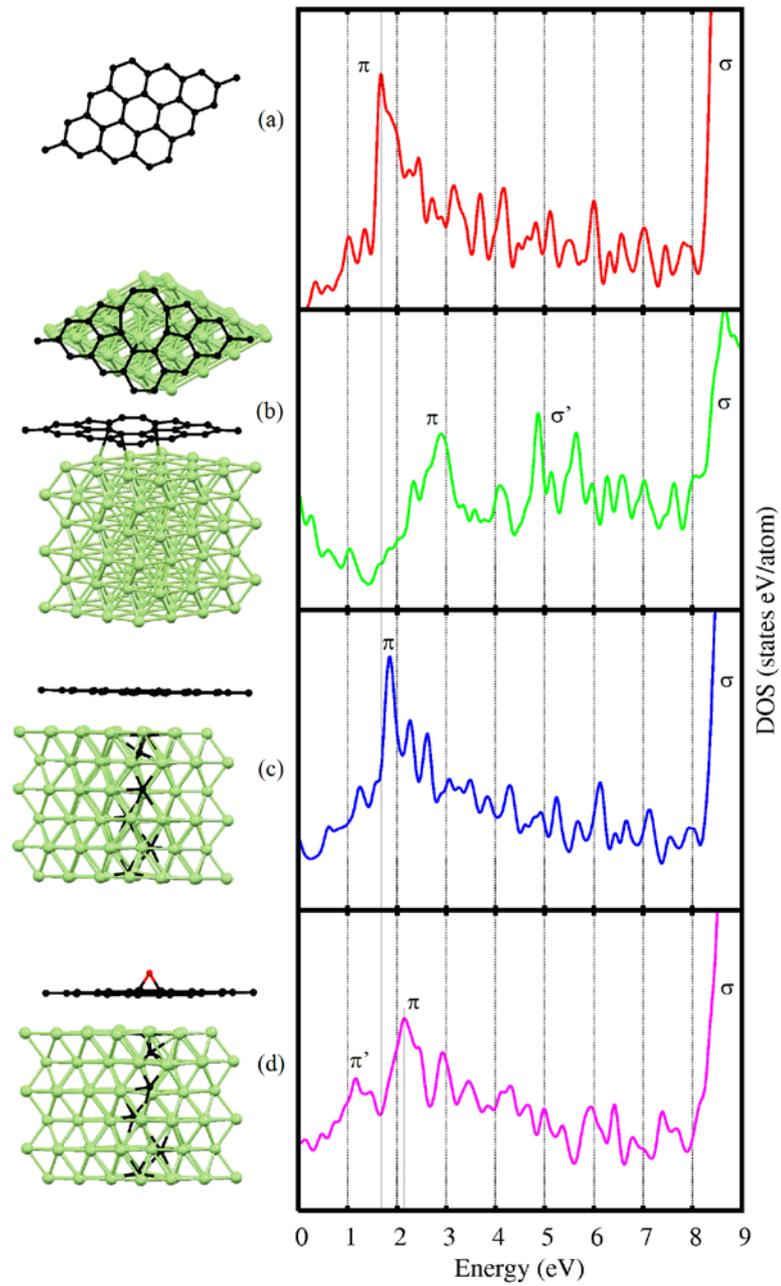

**Figure 5.** Optimized atomic structures and density of the unoccupied carbon 2p-states of the studied types of graphene: (a) free-standing graphene, (b) graphene with bi-vacancies on the Cu substrate, (c) graphene on the Cu-substrate with structural carbon defects, and (d) partially oxidized graphene on the Cu substrate with structural carbon defects.

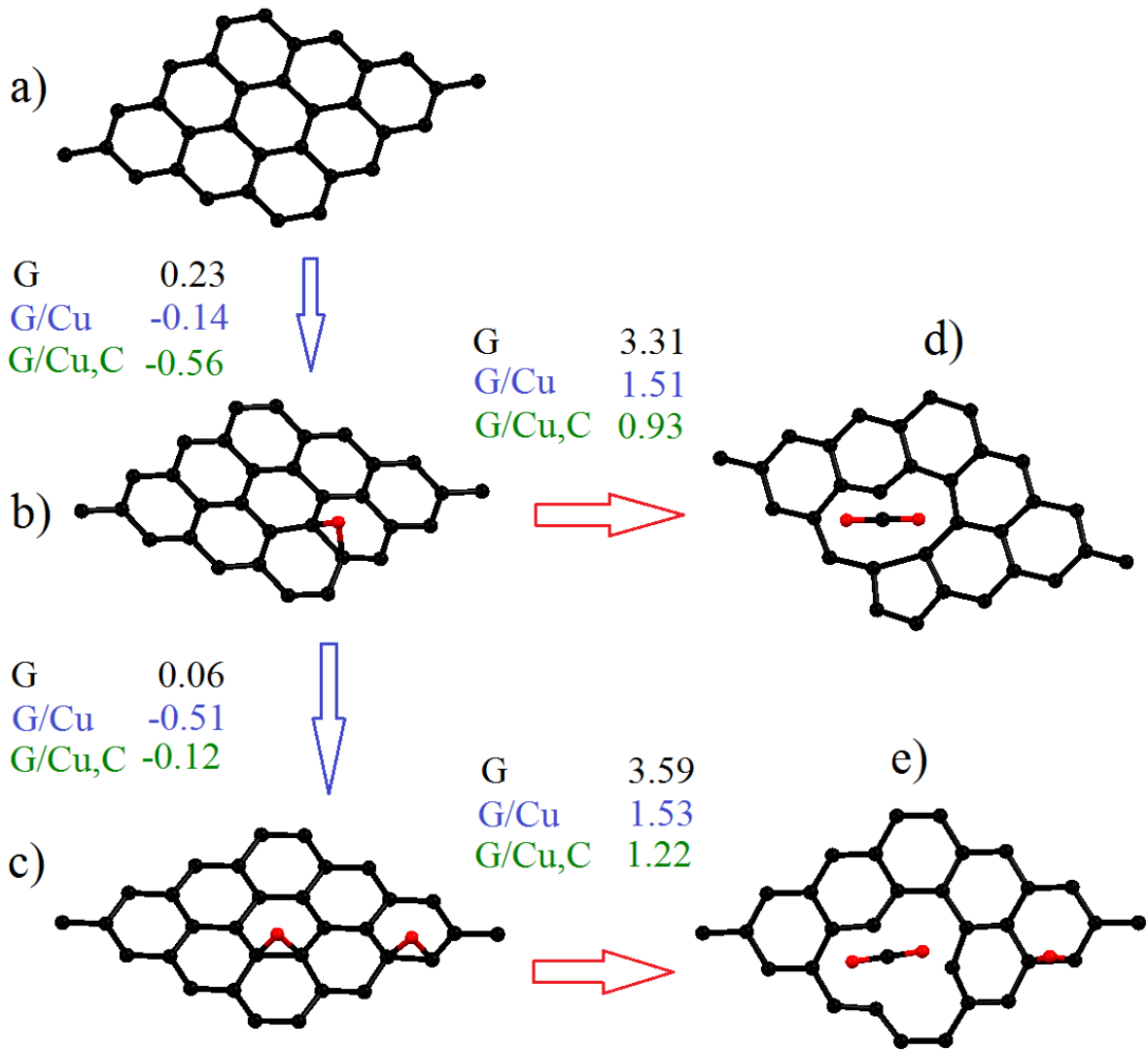

**Figure 6.** Optimized atomic structures of free-standing graphene and formation energies of structural defects (in eV) for pure graphene (Gr), graphene on the Cu-substrate without defects (Gr/Cu), graphene on Cu-substrate with carbon defects (Gr/Cu,C) for oxidation (b and c), and perforation with removal of carbon atoms from oxidized graphene in the form of $CO_2$ (d and e).

The formation of vacancies enables the penetration of oxygen into the copper substrate, which is considered to be a two-step process [26, 33]. In the first step, adsorption of oxygen will occur from the formation of epoxy groups (Fig. 6a,b), followed by the addition of oxygen (Fig. 6c) to form either a second epoxy group or removal of the carbon atoms producing carbon dioxide and a graphene mono-vacancy (Fig. 6d,e). To estimate the influence of the Cu substrate, three different cases are considered: free standing graphene (Gr), graphene on Cu-substrate (Gr/Cu), and graphene on Cu-substrate with embedded carbon (Gr/Cu,C). For free standing graphene, oxidation is an endothermic process, and mono-vacancy requires high formation energy after oxidation (3.31 eV), so free standing graphene is stable towards mono-vacancy formation in air. It has been shown experimentally and theoretically that the presence of a metal substrate reduces the chemisorption energy of various species [19,26]. The same conclusion can be arrived at from our calculations for the case of oxygen because the first step of oxidation is an exothermic process (Fig. 6b,c). It is further found that the Cu substrate not only facilitates graphene oxidation but also reduces the formation energy of vacancies because after the removal of the carbon atom, broken C-C bonds do not become dangling but form covalent bonds with the metal substrate (similar to Fig. 5b). However, the vacancy formation energy is still rather high (about 1.5 eV). Using the previously proposed method of comparing the calculated formation energies and experimental temperatures [44], we can conclude that in the case of weakly oxidized graphene on a Cu substrate, vacancy formation will occur at temperatures around 200 °C. The obtained result is in qualitative agreement with the experimental results of the observation of copper substrate oxidation after annealing of graphene/Cu systems [7,8] or changes in the chemical properties of graphene on metals [11,12].

## 6. Conclusions

In conclusion, we have demonstrated the properties of graphene as a protective layer for Cu. The first-principles DFT calculations show that oxidation does not occur when the graphene monolayer is defect free. In the case of carbon vacancies, the interface of graphene/Cu becomes permeable to the ingress of oxygen atoms, resulting in metal oxidation. The presence of defects in the copper substrate significantly decreases the energy cost of the oxidation and perforation of graphene. We have found from X-ray spectroscopic measurements that there is no evidence of electrochemical corrosion of copper covered by graphene after oxygen exposure at atmospheric pressure from 1 month to 1.5 years. Therefore, perfect graphene without defects or large grain boundaries can preserve the surface of copper from corrosion in ambient air over a long period of time. Another result from experimental measurements and theoretical modeling is observation of a shift in the π* peak in the electronic structure of graphene at 0.57 eV from its position in graphene on $SiO_2$. This causes spontaneous oxidation and the presence of impurities in the metallic substrate.

Based on the results of theoretical and experimental evaluation, we propose that a graphene monolayer is unsuitable for application in industry as an anti-corrosion cover of metallic surfaces because only a high-quality defect-free graphene cover provides sufficient protection of the metallic substrate from oxidation [5]. The presence of defects and impurities, which is unavoidable for large scale industrial production, significantly decreases the protective properties of graphene [7]. The solution could be the turn from coverage by a less chemically stable monolayer to a more stable bilayer graphene [12,33] that does not significantly affect the catalytic [26] and electronic [20] properties of the coper/graphene interface. This possibility requires further detailed examination of the mechanical properties of the interface metals and the graphene bi- and multilayers.

**Acknowledgements**

XPS measurements were supported by the Russian Science Foundation (Project 14-22-00004). D.W.B. acknowledges support from the Ministry of Education and Science of the Russian Federation, Project №3.7372.2017/БЧ. S.O.C. acknowledges the support of the Government of the Russian Federation (Act 211, agreement № 02.A03.21.0006). G.S.C gratefully acknowledges the support of the Natural Sciences and Engineering Research Council of Canada (NSERC) and the Canada Foundation for Innovation (CFI). Research described in this paper was performed at the Canadian Light Source, which is funded by the Canada Foundation for Innovation, the Natural Sciences and Engineering Research Council of Canada, the National Research Council Canada, the Canadian Institutes of Health Research, the Government of Saskatchewan, Western Economic Diversification Canada, and the University of Saskatchewan.